# Darknet Data Mining: A Canadian Cyber-Crime Perspective

Edward Crowder
Institute of Technology and
Advanced Learning
Sheridan College
Oakville, Canada
crowdere@sheridancollege.ca

Jay Lansiquot
Institute of Technology and
Advanced Learning
Sheridan College
Oakville, Canada
lansiquoj@sheridancollege.ca

**Abstract—** Exploring the darknet can be a daunting task; this paper explores the application of data mining the darknet within a Canadian cybercrime perspective. Measuring activity through marketplace analysis and vendor attribution has proven difficult in the past. Observing different aspects of the darknet and implementing methods of monitoring and collecting data in the hopes of connecting contributions to the darknet marketplaces to and from Canada. The significant findings include a small Canadian presence, measured the product categories, and attribution of one cross-marketplace vendor through data visualization. The results were made possible through a multi-stage processing pipeline, including data crawling, scraping, and parsing. The primary future works include enhancing the pipeline to include other media, such as web forums, chatrooms, and emails. Applying machine learning models like natural language processing or sentiment analysis could prove beneficial during investigations.

## I. Introduction

To fully understand the threat landscape, you first must correctly identify and fully understand the threat model of an enterprise or country. This research project explores the darknet and its applications by means of data mining. The results include an analysis of current and past darknet marketplaces, a data model capable of further machine learning for indicators of compromise (IOC) analysis, and value analysis for identifying threats in the darknet. Presenting a sample application that includes a web interface consisting of organized threat information visualized for a qualified analyst to make strategic cyber decisions.

To maintain a common terminology, the darknet is a resource that cannot be accessed without The Onion Router ("TOR") [1, 3]. The Tor Project, a 501(c)3 US nonprofit, advocates human rights and the defense of a users privacy online through free software and open networks[16]. There are many benefits to the darknet, such as online anonymity and enhanced privacy[20]. However, in a recent survey of 25,229 general internet users by the Centre for International Governance Innovation (CIGI) that took place across North America, Latin America, Europe, the Middle East, Africa, and the Asia-Pacific region these same tools perceive an increase to exacerbated cybercrime[19]. The project discusses the risks associated with using the darknet as both a user and cybercrime analyst, followed by an outline of the proposed system design.

Considering benefits and threats that Tor may pose would be beneficial. There was an increase in the Tor network usage, specifically in Canada[16] between 2019 and 2020. At a minimum, this project may uncover why more Canadians turn to the Tor network as shown in Fig. 1., and what the intended usage is.

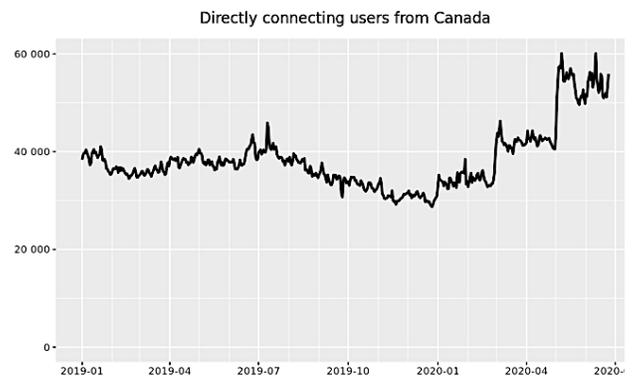

Fig. 1. Tor relay for Canadian users between 2019-01-01 – 2020-06-27 [16]

## II. Related Work

Web crawlers have been around since the early 1990's[10]. The most notable of all web crawler projects being S. Brin and L. Page's Google[11]. As the internet grew, it segregated into multiple layers, known as the Clearnet, Deepnet, and Darknet[12]. Exploring the Darknet provides many benefits if done correctly and in a time-sensitive manner. Researchers have found great value in extracting indicators for use in private companies, government, and personal protection [2, 4, 5, 6]. As a result, many large data sets such as the Darknet Marketplace Archive (DNM) from 2011-2015 [7] are publicly available. However, it was decided not to use the data sets provided for many reasons. First, with only 6 of the 89 DNMs remain accessible[3], it would be a better representation of current technologies used in new darknet marketplaces to data-mine active markets instead of referring to dead sites.

Understanding the potential ethical, moral, and physical risks surrounding the darknet is also essential to keep in mind. Martin et al. [1] explore the significant uncertainties regarding the ethical dimensions of crypto market research. Furthermore, there are so many different environments (e.g., web pages, chat rooms, e-mail) and that there are new ones continually emerging means that explicit [ethical] rules are not possible [1]. Ethical problems are demonstrated by example, through the use of known ethical principals and collaborations with others involved in the study of cryptomarkets. Martin et al. further discuss the risks and threats of assessment, geographical concerns, copyright issues, the effects on the public, and academic research considerations, such as determining national jurisdiction, self-critical awareness of the potential for bias, and many more. Their research concludes with open-ended questions for the researchers to consider metacognition regarding the decisions within their project.

Dittus et al.[4] performed a large scale systematic data collection of the darknet in mid-2017, which claimed to cover 80% of the darknet. Their findings show that 70% of global trades are attributable to the "top five" countries: USA, U.K., Australia, Germany, and Holland. Their research shows Canada falls in sixth place within their findings. Research also suggests that the darknet is not revolutionizing this crime. However, it changes only the "last mile" and only in high consumer countries, leaving old drug trafficking routes still intact.[4, 13]. The publishing of this research topic is a prime example of an ethical dilemma Martin et al. discussed. This paper's results could negatively influence the public funding surrounding the darknet drug trade because of it. The project aims to create a system where information is provided for a qualified analyst to weigh in from their experience and not overstate risks.

Nunes et al. [2] present an operational system for cyber threat intelligence gathering from various sites on the Darknet. Nunes et al. focused on malicious indicators such as threat actor names, private sale of data, and executables that they utilized to fulfill their primary intelligence requirements for emerging threat detection. The creation of a focused web crawler, as opposed to a generic web crawler, was required to collect a vast amount of data. Static processing was done after mass collection to extract indicators of interest. Specializing in cross-site connections, Nunes et al. created a connected graph depicting their indicator attributions to the underground threat actor profiles. Lawrence et al. [3] continue to work in this direction with their product, D-Miner. D-Miner is a darknet "focused" web scraper that collects and parses out specific darknet marketplace features. By utilizing JSON, Lawrence et al. gain the benefits of indexing the data in Elasticsearch. Elasticsearch is a search engine based on the apache lucence search engine. The power of Elasticsearch allows Lawrence et al. to utilize features such as full-text search and REST API's[9]. Data visualization is made possible through Kibana, an open-source data visualization dashboard for Elasticsearch[14].

A primary issue surrounding the Nunes et al. project was the use of anti-scraping technologies deployed on the darknet. The solution proposed was death-by-captcha (DBC), a paid service to solve captcha codes to automate the solution [3]. Hayes et al. [5] take a similar approach to their analysis by identifying the vendors. They explore the use of Apple Script and the Maltego investigation platform to generate a cross-site threat actor connected graph[15]. Notably, the authors outsourced the requirement of solving captcha to the analyst, and the extraction and enrichment process to Maltego's built-in investigation transformers. Combined, this made a robust framework for the manual analyst. However, it was not scalable to the extent desired, and therefore, decided to continue with the use of Python and custom interfacing options.

III. RISK ASSESSMENT

This project uses passive fieldwork, that is, only observing publicly available material that does not require direct communication or response to possible nefarious actors except for account verification. The alternative to this would be active fieldwork, which would include participation within the darknet communities[21].

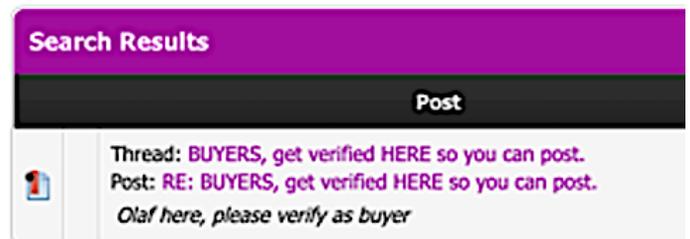

Fig. 2. Active Fieldwork exception for account verification post on darknet marketplace using username "Olaf"

Identity and geolocation protection is made possible by utilizing the Amazons Web Services ("AWS") public cloud as a technical security mechanism. If the crawler, scraper, or other communication to the darknet were to be compromised somehow, due to misconfiguration or other, the researcher's primary machines would remain safe. AWS provided an additional hop from the tor circuits, among many other development benefits. Secondly, one nuance with operating this project within Canada is the law surrounding illicit images found on the darknet. The current minimum penalty for possession of, or "accessing," child sexual exploitation material is six months of imprisonment[1], and the amount of child sexual exploitation material detected online by law enforcement and the private sector continues to increase[22]. The web crawlers and scrapers by design are as focused as possible to prevent crawling outside the scope. A security mechanism to avoid downloading illicit images is made possible by not crawling to URL contained within  tags. This method of risk mitigation avoids the collection of  tags, losing very little analytical value. It altogether prevents the possession of any illicit images that may get us in trouble. All HTML files collected are then parsed, extracting only relevant information and then delete promptly afterward. This aids in allocating more space for new data but, more importantly, protects the user scraping and crawling.

Furthermore, Dittus et al. identified some marketplaces that outline a Terms of Service (ToS) restricting web crawling on their websites similar to legal operation businesses. To make data collection possible, the acknowledgment and accept the risk of violating these terms of service. Arguably, the public marketplaces accessed are available to the public.

## IV. Data Collection

Scraping the darknet is almost the same process as scraping the clear net. The HyperText Markup Language ("HTML") of the page has the contents of interest; it is just a matter of extracting the data in a dynamic and practical format that enables flexible indexing.

The technical options to scrape the contents of pages are broad and it is not a novel function by nature of the internet[11]. There are many frameworks and heavily featured libraries that aid in making the scraping process streamlined and efficient. Options for the use case of scraping the Tor Network[16] included utilizing, Apple Script, Linux utility Curl, and Python coupled with a library Beautiful Soup.

Apple Script is a proprietary scripting language invented by Apple Inc. that aids in the automation of Mac applications[17]. Similar projects tailored to scraping have used Apple Script to automate the process of scraping HTML and partially circumvent security roadblocks such as CAPTCHA. Martin Dittus et al. used Apple Script to automate the bulk of scraping, and when prompted by a CAPTCHA or other automation blockers, it would send the challenge to a human being to manually solve, and the script would continue to function[5]. Apple Script's utility is compelling; however, due to the proprietary nature (i.e., being locked into using MacOS) and lack of scalability, it did not fall in the appropriate use case for this project.

cURL is a command-line utility that supports a comprehensive set of protocols and is the backbone for numerous applications. cURL, coupled with Python and Bash, presented itself as another avenue to take to solve the problem of scraping the darknet. With its robust feature set, speed, and extensive documentation, the utility showed promise. However, after attempting early builds of the scraper, the technology was not modular enough for the needs of scraping multiple platforms without a substantial restructuring of the supporting codebase, which in return became too time-consuming to continue developing with it.

Python is an interpreted, high-level programming language[18] combined with specific libraries such as BeautifulSoup and Requests. It allows for handling HTTP requests in a format easy to read and use. Lawrence et al. used Python extensively while building a framework to crawl and scrape the darknet, from a practical standpoint, Python was the ideal technology that could get integrated into the stack and be the backbone of the crawler and scraper.

The supporting stack included utilities such as cron, a time-based job scheduler for Unix operating systems. Cron is essential to the project as it helps automate the crawling and scraping of the marketplaces. Rsync, a utility for transferring and synchronizing files between the directories and remote computers, was essential for relocating files, compressing the files at their destination, and backing up to a remote computer.

Every marketplace poses its own unique challenges when tackling the issues of crawling and scraping. Security mechanisms such as CAPTCHA, valid sessions, random URL IDs, and rate-limiting were all challenges met while designing and testing the application. The approach to circumvent CAPTCHA and to acquire a valid session required human interaction at the beginning. If the marketplace utilized encoded or random URL IDs to make it harder to iterate, the crawler would navigate its way to find every link on the marketplace instead, which would add a significant amount of time to the process. Section IX discusses the implementation of a CAPTCHA solver.

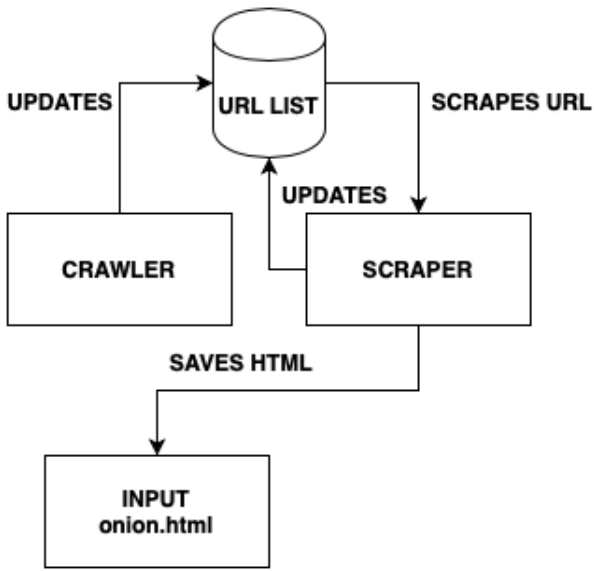

**Fig. 3. Darknet Scraper and Crawler Overview**

Rate limiting was the next hurdle. Every site on the internet handles this differently and the same is true for darknet marketplaces. Some sites would return a 429 Request (Too Many Requests) appropriately when requests made to fast while crawling or scraping. Some marketplaces failed silently, and a small portion did not have any rate limiting whatsoever. Rate limiting is a more trivial problem to deal with, figuring out the rate in which the crawler and scraper to run is simply a matter of trial and error. Eventually, the frequency at which the marketplace processes the requests are tuned to match that of a regular user, therefore circumventing the issue altogether.

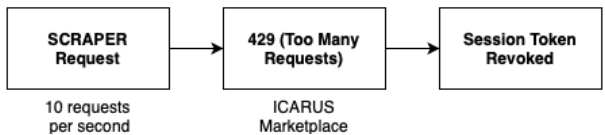

**Fig. 4: Flow of a failed attempt at scraping**

The goal of scraping was to collect an inventory of all items listed for sale. These pages generally have to convey some relationship with the buyer so that it would list information such as base price, quantity, type of item, category, shipping from and shipping to, as well as some more verbose information such as the vendor name, amount of items sold, last logged in time and more. Although this is not consistent across every marketplace, it enriches the data collected, and its usefulness to tell a story is further discussed in section VI.

```
Algorithm 1. Crawling Page
for each URL in page
    save onion_link to queue

Algorithm 2. Queue
    for each URL in queue
        if URL == 200_OK
            add to active_links
        else
            delete from queue

Algorithm 3. Scrape
    for each URL in active_links
        scrape HTML of page
        save file to scraped_links as name of onion_link
        timestamp

Algorithm 4. Parse
    for each file in scraped_links
        retreieve information matching json object
        save retrieved information as json object
```

**Fig. 5. Psuedo code of data collection and parsing**

## V. DATA PROCESSING

After the collection process, the collector has provided HTML files containing the original structure from the advertisements on the target marketplaces. Most of the residue data is qualitative such as usernames, countries, product categories, and many more; however, it is also possible to extract quantitative data, such as sales, review count, view count per post, and post frequency by vendors. The following paragraphs will describe an overview of the data processing pipeline and considerations regarding data storage, data consistency/redundancy, data availability, and data scalability. It will discuss the choices made and the processes applied at each step to arrive at the final solution.

As described in previous sections, the crawler and scrapper push their files to an inbox, and the data parsers pick them up and push to an outbox, and Elasticsearch database. The primary reason to delegate particular tasks to each part of the process was intentional to avoid bloating one piece of software with many features. The decoupling of responsibility provides the ability to change and modify code with ease since they are merely new class files based on parent objects with slight modifications to fit the varied requirements of many marketplaces.

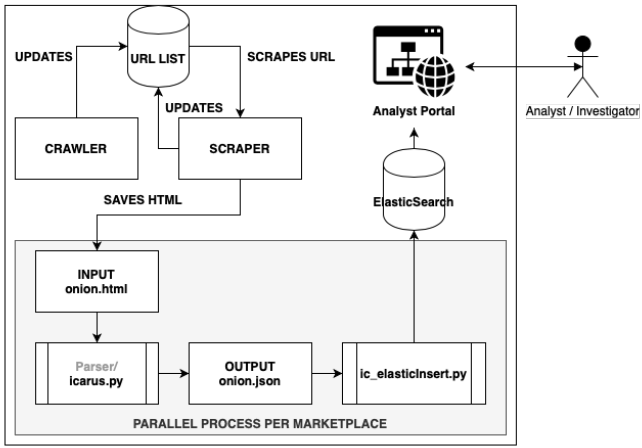

Fig. 6: Darknet Stack System Design Overview

The First requirement examined was data storage. Data storage was a significant consideration because of the project's nature generating large amounts of files from the project's inception. As more and more files in a directory accumulate, it can become quite costly due to the read/write overhead compared to a single larger file. Maintaining the original HTML files' integrity was made possible because of these considerations. However, the entire file does not benefit the end goal of the analysis. The data processor creates JSON objects from the original HTML files by extracting only the features required for analysis. This process saw a reduction of 167% on average **Fig. 7.**, yielding only the features expected to work with; the new JSON file is from here on out referred to as the Darknet Data Object (DNDO). With this new DNDO providing the full flexibility of JSON Elasticsearch can utilize some powerful indexing features that predefine data types for storage optimization. The decision to store the file locally during testing and index it into Elasticsearch 7.8 is possible but consumes more storage, which is discussed in further detail later.

| MARKETPLACE | HTML AVG | DNDO AVG | REDUCTION |
|---|---|---|---|
| Elite market | 5.06 | 1.23 | 121.79% |
| Icarus | 14.37 | 0.61 | 183.52% |
| Aesan | 76.05 | 0.68 | 196.42% |
| **TOTAL** | | | **167.24%** |

Fig 7: HTML to JSON size reduction findings in KB

Second, the issue of data consistency and data redundancy. With every darknet marketplace containing different structures and data layouts, it was essential to consider data consistency for the cross-marketplace comparison. Without consistent data features, the cross-marketplace quantitative analysis would not be possible. The significant benefits of switching from HTML to JSON were extracting and removing all HTML structure, CSS and JavaScript imports, images, and repeating headers/footers. Reducing the redundancy and overhead allows the information to remain consistent across all marketplaces and control values that are missing with expected values. It is also possible to extend the DNDO at any time and maintain previous analytics. Overall, the flexibility offered creates a lower redundancy and higher consistency.

The third part of processing pipeline requirements includes the availability of the data. The entire lifespan, from collection to full-text search, and even data expiration, ensured the data was accessible in all forms. Managing data in lifecycles like this provided the freedom to experiment without concern about the loss of data. The worst-case scenario was re-processing and indexing the original HTML files, which at scale did not take more than a few minutes. However, paired with availability comes scalability. The decision to choose JSON was to ensure that the chosen format can adapt to any need during this project's future work. The primary concern in this vector was to avoid redundancy and index what only the required information required to ensure the data was always fast to query. By utilizing Elasticsearch's ability to predefine the data types of incoming JSON files, it is possible to tune and maintain this speed compared to ingesting raw text.

| CONSIDERATIONS | INDEX HTML | INDEX DNDO |
|---|---|---|
| Storage Requirements | High | Low |
| Data Consistency | Low | High |
| Data Redundancy | High | Low |
| Data Availability | Low | High |
| Data Scalability | Low | High |

Fig 8: Table of data considerations

Finally, after careful consideration, the benefits and downsides of extracting information from the raw HTML and converting into specific focused, and extendable JSON objects that are smaller and more efficiently handle the issues of data redundancy and consistency. JSON allows complete freedom to extend the object at any time and maintain previous analysis. The plethora of libraries and tools available to work with JSON make it is an obvious choice. By design, high process interoperability allows for a flexible shift and switch of the backend system should future marketplace analysis require that. Breaking up the system components into specific small services per marketplace allows the process to scale horizontally and vertically in the cloud with ease. Due to the nature of the process, cross-cloud interconnectivity is also possible so long as every cloud, private, public, or hybrid has access to the same Elasticsearch cluster. Furthermore, it is possible to shift focus entirely and modify the DNDO to meet a specific industry need, altogether avoiding marketplaces if required.

## VI. DATA ANALYSIS

Throughout all the marketplaces we analyzed, the old saying of "Honor Among Thieves" stands true to darknet markets.

Of all the marketplaces scraped, each had strict rules on the sale of guns, child sexual exploitation material, and murder for hire. Each marketplace observed specific one category, which strengthens their reputation among users.

The primary goal of data analysis was to validate the output of the data collection process. The secondary goal of data collection and analysis was to set the stage for larger projects such as machine learning and predictive analysis. This section outlines the data analysis, specifically text analysis, statistical analysis (What happened), and diagnostic analysis (Why did it happen). Afterward, the data interpretation process involves the decision making behind how we chose to express our results and the groundwork for visualization. Finally, Kibana and the custom data interface are the tools of choice to provide the data visualization methods used on the current data sets.

To make the most of the data, there was a high focus on the fields which contained non-null values across every darknet marketplace. The first field identified that met these criteria was productClass. The productClass field contained two values, physical or digital. Classifying the marketplace posting based on the end product form factor made sense to start with since the two separate categories have very different risk profiles. Approximately 66% of all products collected were Digital compared to the 33% physical items. Further, the data collected identified that most marketplaces are highly targeted and tailored to specific categories of products.

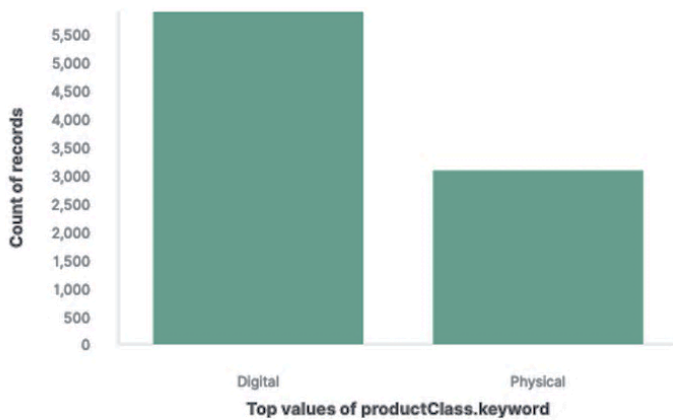

**Fig. 9. Product Class type Digital (left) vs. Physical (right) products**

The ASEAN market data shows mostly digital goods instead of physical. However, EliteMarket shows the opposite. Interpreting the data collected for the ASEAN Market shows that the marketplace primarily focuses on digital goods. The top 5 users control 38% of the market posts, and 82% of the shipments can be worldwide. By creating a heatmap of data for ASEAN, it was clear that the market specialized in digital products. The Y-axis shows the top 5 categories, and the X-axis shows the origin country of shipping. As depicted in Fig 10. most of the products are listed worldwide, France, the United States, and Germany.

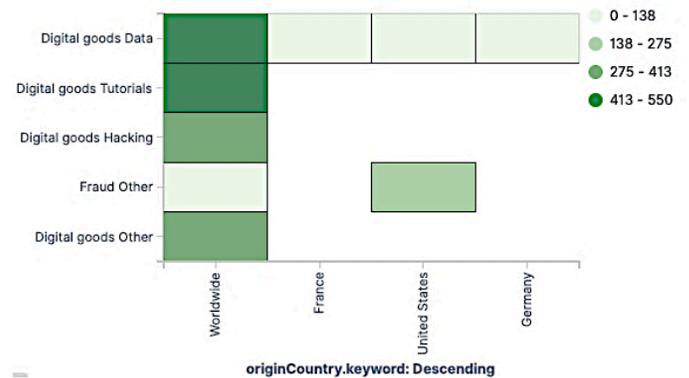

**Fig. 10: ASEAN Category heatmap**

Elitemarket shows similar statistics with their shipping being 95% worldwide for both origin and destination country; however, their focus is primarily on physical goods. The first concerning piece of information extracted is that the primary origin country is worldwide. The use of worldwide as an origin country is concerning for many reasons; firstly, the vendor could be concealing their origin country by listing their product origin country as worldwide instead of their actual destination. Secondly, the vendor may be truthful, which means their illicit drug supply chain is truly global, which makes apprehension difficult for any authority.

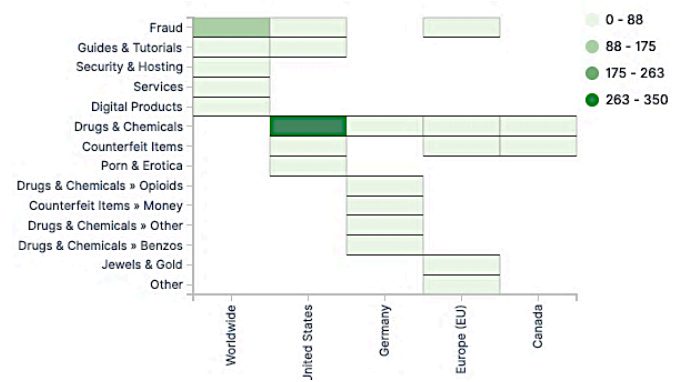

**Fig. 11. EliteMarket Category heatmap**

Fig 11. shows Canada's primary exports as "Drugs and Chemicals" and "Counterfit items." The RCMP Supplementary Estimates for the 2019-2020 budget predicts only 15 Million Canadian dollars set aside for fighting cybercrime of all types, including the prevention of illicit drugs[25].

It is possible that a cross-market seller, DrunkDragon exists. DrunkDragon appears to sell 48% of all digital products collected shown in Fig. 12. Although, there is no way to validate this, as it could merely be an imposter trying to build immediate trust by using a well-known name.

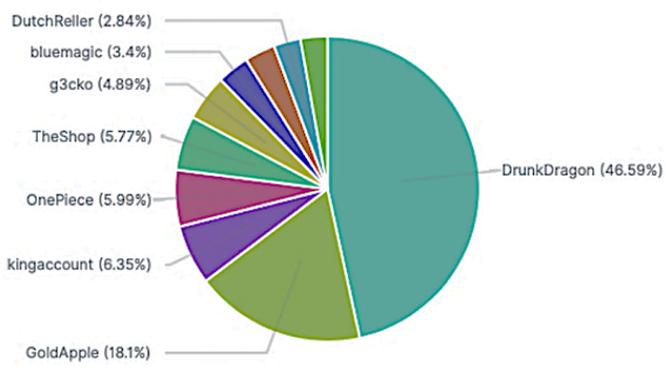

**Fig. 12. Top Sellers globally**

The darknet operates on a near-zero trust model. Every transaction observed required escrow services. Escrow is a trusted third party that holds funds and or the item until the transaction has been completed by both parties, as shown in **Fig. 13**; this is standard procedure whenever an item is to be traded or sold on any darknet marketplace recorded. This service is usually pay-per-transaction exclusively using cryptocurrency; it is either added to the item's total cost or handled entirely separate from the initial transaction.

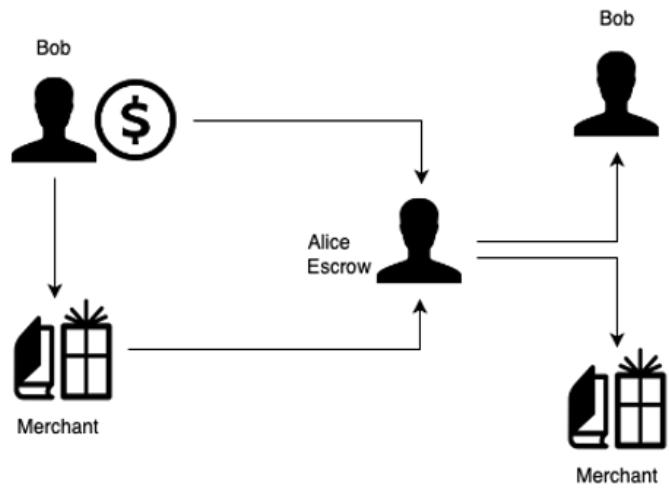

**Fig. 13. Escrow Service flow**

A cryptocurrency tumbler, also known as a mixing service, is typically used in the escrow process. It allows the user to pass the bitcoin funds from one address to a chain of random bitcoin addresses and then finally to a new address in control of the user. The randomization of bitcoin transfers supposedly improves the anonymity of cryptocurrencies that are not as privacy-focused, such as Bitcoin. Cryptocurrencies such as Monero (MNR) and Zcash are popular alternatives to the popular cryptocurrency Bitcoin. They provide the anonymity of transactions needed for these marketplaces to flourish. Monero achieves this by using an obfuscated public ledger, which allows anyone to receive or send transactions, but no single user or entity can observe the source, destination, or amount. Hence why most marketplaces have been moving to privacy-focused forms of cryptocurrency to protect the buyer, seller, and escrow.

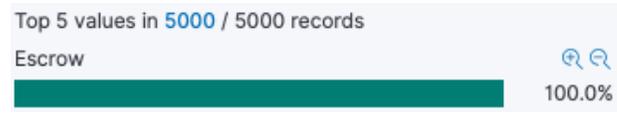

**Fig. 14. 100% of sales/trades via Escrow**

There is very little price disparity of digital goods on darknet market places due to an unlimited supply of the digital items sold. Some of the most common digital items seen include PDFS and video guides. The topics of these documents and videos range from social engineering free products from companies to illicit drug production at home. The sheer abundance of guides on the same topics dilutes the market and drives prices to below 10 US Dollars (USD), as displayed in **Fig. 15**. The supply versus demand makes the information extremely available to even those with limited funds.

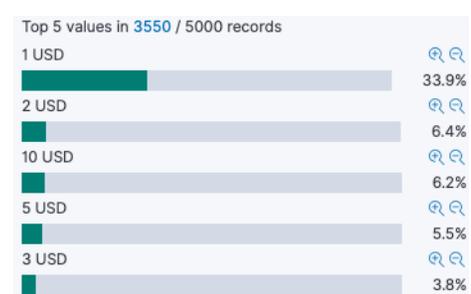

**Fig. 15. Price differences of digital goods**

Quantities of items is a section almost every marketplace vendor has to advertise publicly, the most typical amount seen from the data collected is "Unlimited" or the max amount a vendor can enter in the field usually "999" or "99". Maximizing the quantity works in the vendor's favor to help disguise the amount of the product the vendor owns. It is also more convenient for the vendor to enter the maximum value than the actual amount as it appears they have a more extensive operation, and it is easier to set and ignore compared to updating stock. Therefore, this key and value pair is rendered useless from a data visualization standpoint as it gives almost no insight into what a vendor has in stock, as shown in **Fig. 16**.

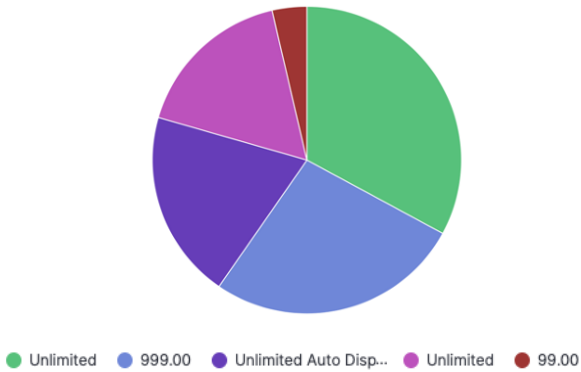

**Fig. 16. Top 5 Quantities across marketplaces**

VII. EXTERNAL INTERFACE PLATFORM POC

Although Kibana provides an intuitive and transparent way to display, search, and visualize the data collected, it cannot modify stored data. Modifying the fields prepended with "analyst_" in the DNDO allows the user to track their research and thoughts on each record. The lack of data modification is why it was required to go beyond the ElasticStack. Like Kibana, it is possible to leverage Elasticsearch's extensive API. It is possible to use a platform like Flask and the open-source Elasticsearch client python package to create a custom third party interface. Flask is a lightweight Web Server Gateway Interface (WSGI) web application framework designed to make getting started quickly and easily, with the ability to scale up to complex applications[23]. The Elasticsearch python library provides a wrapper to connect, create, update, and delete records quickly.

Introducing FlaskDash, a Python-based web application for analysts and researchers to manage, flag, save notes, and create new insight into darknet marketplace analysis. In its current state, the platform is incomplete. However, it provides a base for future work to be extended upon in many ways. The groundwork for this platform and the minimum viable product was to host an instance of the Flask WGSI connecting to Elsaticsearach while supporting multiple index and record navigation.

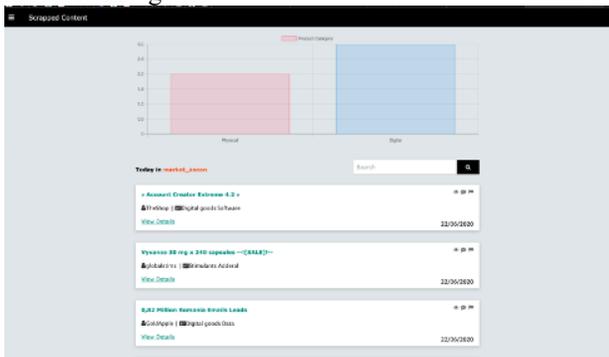

**Fig. 17. FlaskDash Initial landing page overivew**

The ability to search for records was critical to the success of this application. The Elasticsearch connector made it possible to query the database for any of the provided fields. The "title" field provided the ability to search for common words across many records. Elasticsearch's ability to analyze the English language made it possible to search for a specific search term such as "Account" and return variations, such as the pluralization of that word. It is also possible to configure filters to search on any field available in the DNDO, or search across the entire message if unsure which field.

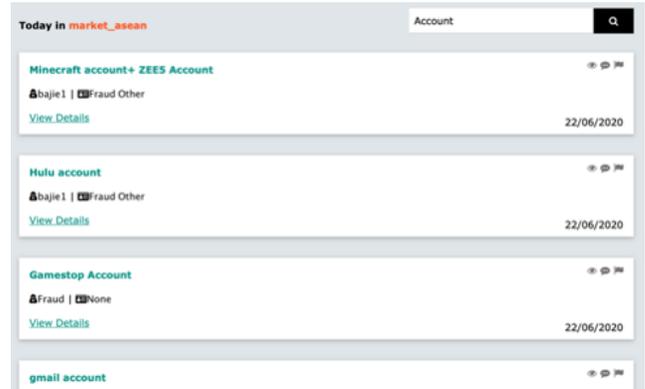

**Fig. 18. FlaskDash search functionality**

The three buttons in the top right corner of every record are visual cues for quick status updates. These buttons include anayst_viewed, analyst_comments, and analyst_flagged. The analyst_viewed field is a boolean data type designed to identify if an analyst has quickly viewed the record. If the analyst_viewed value is false, the button will remain grey, indicating the record is new or unread. The analyst_comment field data type is raw text. There is no upper limit on this field. However, by default, there is a built-in limit in the HTTP (chunk handling) layer that limits requests to 100mb[24]. By leaving notes and analysis, other analysts can review, or search on these notes to correlate a specific case if a case ID is left there. The final button is analyst_flagged and is used to identify records of importance quickly. A record should be flagged if it required further analysis or is under current investigation. However, if these fields do not meet every team's needs, the JSON fields can be extended without corrupting the previous data, and new buttons created.

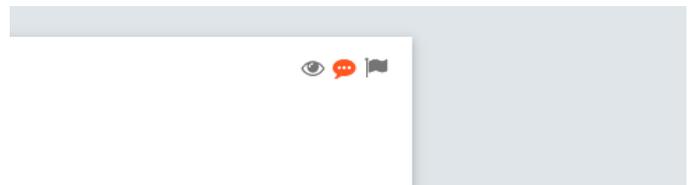

**Fig. 19. FlaskDash analyst buttons**

Although Kibana is a fully functioning dashboard and analyst platform itself, it cannot modify the data directly. FlaskDash, at this point, meets the minimal working product requirements, and there is much to be done discussed later in future works. The goal of this platform was to test the viability of viewing DNDO in an external interface and creating an additional hop between the analyst and the darknet itself. Retaining the same abilities to search and view the records was successful, and having this air-gap protects the end analyst and their machine from the direct risks of browsing the marketplace themselves.

## VIII. CONCLUSIONS

In conclusion, the crawler, scraper, parser, and web application enabled the analysis of current and past marketplaces. The creation of a JSON based Darknet Data Object encompassed features identified from cross-market analysis. The features extracted allowed the data visualization to display the real value of identifying emerging threats within the Darknet.

During the initial research, previous marketplace data provided by the GWERN Darknet Archives assisted in the organization of future data as well as templating valuable information within darknet marketplaces. However, the previous data proved to be of low analysis value due to changes in technology. Primarily, the intelligence value had expired and no longer provided useful context. Current darknet marketplaces offered a surprisingly consistent set of information. This information became the basis for the darknet data object. Having a scalable, consistent set of data allowed for easy cross-market analysis and targeted analysis of single marketplaces. A Canadian presence throughout every market was constant. To this extent, the first goal of the project was successful.
Physical products appeared to be more prevalent within Canada, although digital products made up approximately 66% of the 10,000 records collected. A significant issue with product attribution toward a specific country was the use of "worldwide" shipping. Most marketplaces contained about 2-3% Canadian origin and 90% worldwide, leaving a range of 2-90% Canadian origin possible. The dramatic difference left low confidence in inaccurate attribution.

When targeting vendors, the vendor DrunkDragon, claimed 46.59% of all sales (mostly digital). Looking at Canada only, AtomikBomB sells 70% of all physical products in Canada. However, AtomikBomB is not in the global top 10, attesting to the lack of a Canadian vendor presence within the Darknet. The gap between supply and demand opens up a potentially increased risk for Canadians participating in the exchange of illegal goods.

Overall, the Darknet data object provided valuable in many ways. Building upon data consistency, availability, scalability, and removing redundancy, the cost per query and storage access decreased. The findings within the analysis provide a useful context for Darknet operations. This context, paired with a qualified analyst, can be used to make strategic decisions. The open-source nature and low cost of entry into data mining the Darknet prove valuable to any organization wishing to venture into this landscape. Extending the technology stack with an external web application provides the opportunity to meet the needs of many. The Darknet, at first glance, appeared to be an endless pit of data. However, with the right tooling, the correct data extracted and visualized can tell an exciting story about what exists below the surface.

## IX. FUTURE WORK

Many features had been removed off the list during development due to a lack of time. It was essential to focus on primary functionality instead of the quality of life features. Since the project is result-driven, the sacrifice of some automated functionality in place of results built a substantial backlog of potential system extensions.

A more intelligent parser that allows the user to remove duplicates and modify values in the key pair of the JSON object via Regular expression (RegEX) would significantly improve the accuracy and quality of the data parsed before it gets digested by the data visualizer. The creation of a more dynamic parser that loads configuration files would be significantly beneficial. The ability to write REGEX expressions to a configuration file to extract fields for the DNDO would save a considerable amount of time writing custom parsers for each media. A more extensive parser would be required when expanding to new media types.

Automating the end-to-end flow of the pipeline would tremendously help when trying to visualize the data. As it stands, an analyst must manually provide the workflow with a registered account on the marketplace, and a valid session token. Although an analyst will most likely be required to register the account, the system should obtain a valid session token and bypass the login pages itself in the future. The rest of the pipeline from the crawling, scraping, and visualization can and should be automated to assure data accuracy.

CAPTCHA completion without human intervention via proprietary software such as DeathByCaptcha would significantly improve rate limiting issues met when working against rate limiting on various marketplaces. Having DeathByCaptcha also allows for faster scraping and crawling since most darknet market places prompt users with Captchas after accessing a certain amount of posts, or the rate of requests exceeds the limit set by the marketplace. Ultimately, matching the requests of a standard user slows the crawling process. However, to avoid delaying the crawling process, it is possible to multi-thread the application with multiple Tor circuits. By utilizing various Tor circuits, it would appear as if numerous unique users are visiting the marketplace as per usual.

A goal of this project if the time allotted was to have a completed analyst view, which would allow an analyst to triage, flag, and comment on items and events of interest based on data and keywords powered by Dark Net Data Objects (DNDO) fed into the ELK stack. However, it lacks the feature set and functionality that initially envisioned for it. As it stands, the current investigation platform is a bare-bones example of what is possible. The first significant feature to get working is the analyst buttons. The current user interfaces support this functionality, but the appropriate routes to post data to Elasticsearch has yet to be completed. Enabling pagination to enhance the interface and searchability of the data is also a critical missing piece. Allowing user registration with role-based access control (RBAC) and Single sign-on (SSO) would make the platform one step closer to enterprise-ready. User account control would also enable us to continue work on the flagging and commenting sections to promote collaboration. Adding a separate case management page to identify records in progress would improve workflow and usability. Finally, enabling an enterprise license on Elasticsearch and configuring machine learning and Natural Language Processing would allow users to scrape marketplaces of all languages, translate and identify trends and anomalies. NLP would also open up opportunities to scrape new information outlets such as chatrooms, forums, and websites.

Visualizing the crawler and parser trails could also prove useful—relationship visualization with technology such as a graph database like Neo4J may provide a unique insight into the structure of the darknet. Graph databases could further enhance cross-marketplace analysis by linking adversary pseudo names to other markets.

## Appendix A

### CALCULATION OF FILE SIZE REDUCTION

In order to calculate the file size reduction consistently we utilized some online tools that where modeled after the following equation:

$$\frac{|V_1 - V_2|}{\left[\frac{(V_1 + V_2)}{2}\right]} \times 100 = ?$$

The calculator is available online here. Values v1 and v2 were obtained with the following bash command inside the html folder and json folder respectively in order to collect the average file size of each file. The command works by continuously summing the filezie column to produce an average of the sums.

*find ./ -ls | awk '{sum += $7; n++;} END {print sum/n;}'*

```
~/Development/python/capstone/Data/input/asean master*
) find ./ -ls | awk '{sum += $7; n++;} END {print sum/n;}'
76058.9
```

### SORTING FILES

Accurately pulling URLs that were associated with items in the marketplace required the linux utility grep to output all URLs that were between a certain to characters in length.

*grep -E '^.{114,120}$' infile*

## Appendix B

### Data Analysis: ASEAN Market

Table representing the top 5 sellers controlling 37% of market posts collected

| seller.keyword: Descending | Count |
|---|---|
| DrunkDragon | 1,114 |
| GoldApple | 362 |
| OnePiece | 268 |
| TheShop | 258 |
| PMS | 123 |

Figure showing the worldwide shipping destination

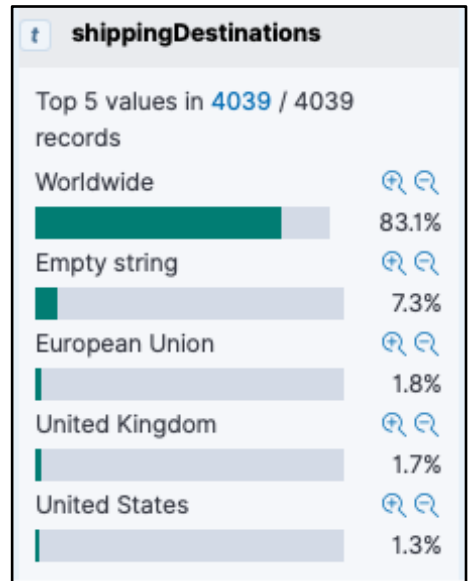

## Appendix C

### Escrow, Pricing, and Quanitiy Issues

100% payment escrow – very good escrow, compare to banks escrow etc..

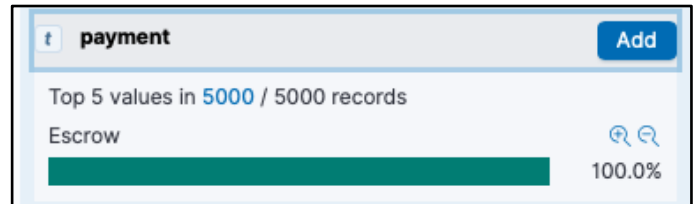

Most digital items are so cheap because easy to reproduce

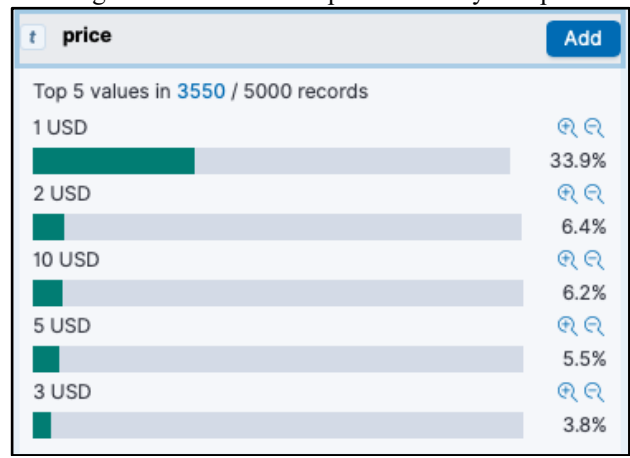

# Appendix D

## FlaskDash investigation platform

FlaskDash initial screen displaying some newly ingested

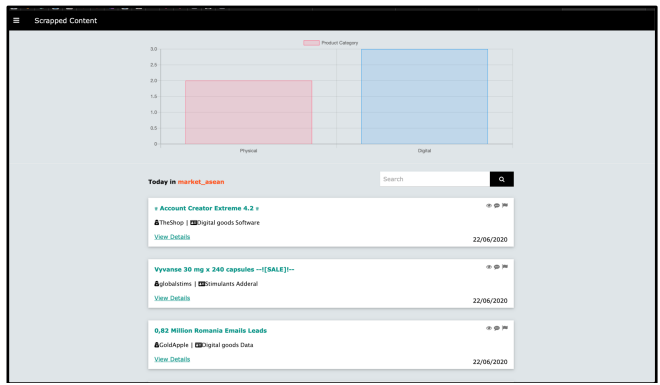

FlaskDash supports multiple elastic indexes via a dropdown menu in its interface.

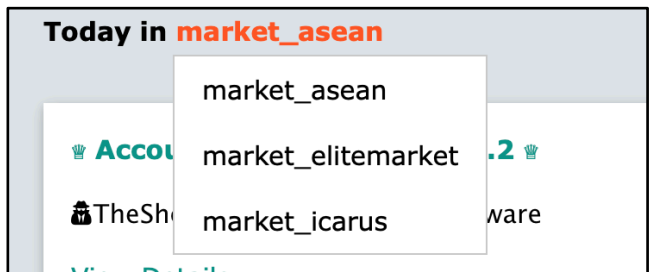

Analyst Viewed, Comments, Flagged fields represented as icons for quicker identification of already viewed alerts displaying the color identification of a completed field.

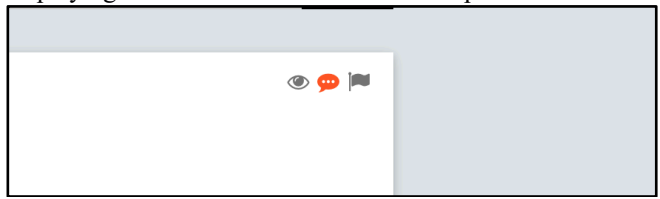

Comments interface modal popup after clicking the item for the respective record. This interface provides a text area and submit button to perform an HTTP Post to the ElasticSearch database.

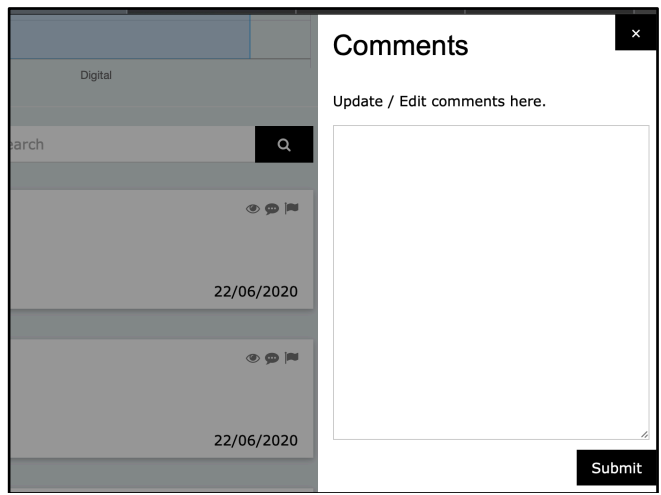

Example of the FlaskDash meta data inside DNDO

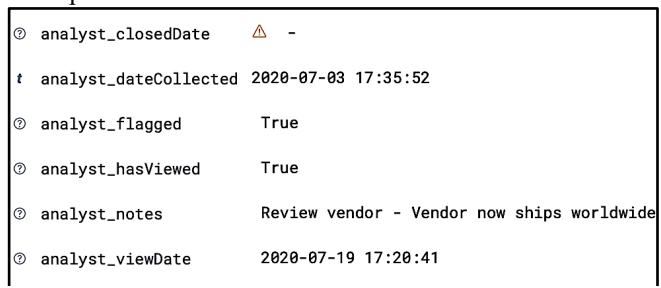

The search funcationality of flaskDash using search terms "account" in the index "Market_asean"

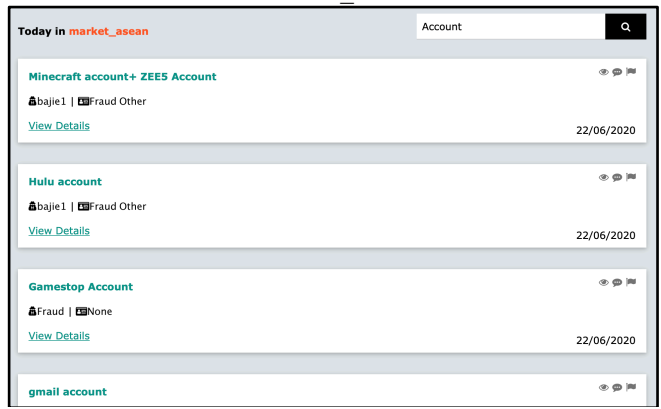

# Appendix E
## Code Resources

**All code resources required to deuplicate the project can be found here:  https://github.com/n0tj/Darknet-Stack**

# Appendix F

## Miscellaneous items

Example JSON dark net data object populated with values extracted from asean market.

```json
{
    "title": "How To Find Cardable Shops ",
    "seller": "DrunkDragon",
    "category": "Digital goods Tutorials",
    "creationDate": "None",
    "url": "httpaseanma4v4iiptqavzdb7endz2bl6es66goytxqow6yk3dmmxyjazfqdonionlisting0a3fcf15bfd64623bb71700b4de592b0.html",
    "views": "None",
    "purchases": "None",
    "expire": "None",
    "productClass": "Digital",
    "originCountry": "Worldwide",
    "shippingDestinations": "Worldwide",
    "quantity": "999.00",
    "payment": "Escrow",
    "price": "1 USD",
    "analyst_hasViewed": null,
    "analyst_viewDate": null,
    "analyst_flagged": null,
    "analyst_notes": null,
    "analyst_closedDate": null,
    "analyst_dateCollected": "2020-07-03 16:56:42"
}
```

Python Code example of Darknet Data Object

```python
# Filename:     DNDO.py
# Description:  Dark Net Data Object
# (DNDO) - Like features across markets for
# ES indexing
# Version:      1
# Date:         June 29 2020
# Author:       Edward Crowder + Jay Lansiquot

class post:
    def __init__(self):
        self.title = None
        self.seller = None
        self.category = None
        self.creationDate = None
        self.url = None
        self.views = None
        self.purchases = None
        self.expire = None
        self.productClass = None
        self.originCountry = None
        self.shippingDestinations = None
        self.quantity = None
        self.payment = None
        self.price = None
        #Non scrapped data
        self.analyst_hasViewed = None
        self.analyst_viewDate = None
        self.analyst_flagged = None
        self.analyst_notes = None
        self.analyst_closedDate = None
        self.analyst_dateCollected = None

    def Post(self, title, seller, category, creationDate, url, views, purchases, expire, productClass, originCountry, shippingDestinations, quantity, price, payment, analyst_hasViewed, analyst_viewDate, analyst_flagged, analyst_notes, analyst_closedDate, analyst_dateCollected):
        self.title = title
        self.seller = seller
        self.category = category
        self.creationDate = creationDate
        self.url = url
        self.views = views
        self.purchases = purchases
        self.expire = expire
        self.productClass = productClass
        self.originCountry = originCountry
        self.shippingDestinations = shippingDestinations
        self.quantity = quantity
        self.price = price
        self.payment = payment
        #None scrapped data
        self.analyst_hasViewed
        self.analyst_flagged
        self.analyst_notes
        self.analyst_viewDate
        self.analyst_closedDate
        self.analyst_dateCollected = analyst_dateCollected

    def toDict(self):
        print(self.__dict__)
```